# Comments on the momentum density and the spatial form of the density-matrix of the Hooke's atom

Sébastien RAGOT

Laboratoire Structure, Propriété et Modélisation des Solides (CNRS, Unité Mixte de Recherche 85-80). École Centrale Paris, Grande Voie des Vignes, 92295 CHATENAY-MALABRY, FRANCE

## Abstract

In a recent paper, A. Akbari, N. H. March and A. Rubio [Phys. Rev. A. 76, 032510 (2007)] have investigated the one-electron reduced density-matrix and the momentum density of several two-electron model atoms, including the Hooke's atom. The method used by the authors for deriving an integral form of the momentum density is well suited for deriving a closed-form expression of the exact reciprocal form factor, which function is of importance inasmuch as it reflects the off-diagonal side of the exact reduced density-matrix.

In a recent paper, A. Akbari, N. H. March and A. Rubio [Phys. Rev. A. 76, 032510 (2007)] have notably investigated the one-electron reduced density-matrix and the momentum density of several two-electron model atoms. Said model atoms include amongst others the Hooke's atom, wherein two electrons are harmonically trapped about a nucleus but repel each others via a Coulomb term.

In short, this model atom was introduced years ago by Kestner and Sinanoglu [1], willing to consider a model *realistic as far as electron correlation is concerned* and for which an exact solution can be formulated, at least in a power series. Later, Kais *et al.* have realized that the exact ground-state wave function could be written in closed-form in the special case of $k = 1/4$ [2]. The subsequent model has therefore logically attracted much attention, offering a springboard to a number of exact properties of confined and correlated electrons [3,4,5,6,7,8].

Besides the wavefunction in position space, a number of properties can be obtained in closed form (see a summary in [3]). However, the fact is that the exact, spinless, one-electron reduced density-matrix $\rho(\mathbf{r},\mathbf{r}')$ can manifestly not be obtained in closed form [3], in spite of interesting attempts to simplify its structure [5,8]. More generally, functions related to the off-diagonal part of the density-matrix (e.g. the reciprocal form factor, momentum density, or the Compton profile) are not known in closed-form. So far, such functions have not drawn much attention, besides the work of March and co-workers (see e.g. ref. [5 - 8]).

In this regards, an integral form of the momentum-density has been proposed in the paper at stake, ref. [8], namely

$$\rho(p) = \qquad\qquad\qquad\qquad\qquad\qquad\qquad\qquad\qquad\qquad\qquad (1)$$

$$\frac{\pi}{8+\sqrt{5\pi}}$$

$$\int_0^\infty db\, \frac{\sin(p\,b)}{p}\, e^{-\frac{3b^2}{16}}$$

$$\left( 8b(b+4) - 64\pi\left(\mathrm{erf}\!\left(\frac{b}{4}\right)-1\right)\mathrm{erf}\!\left(\frac{b}{4}\right)e^{\frac{b^2}{8}} + \right.$$

$$\left. \left(b(8-b^2) + 2(b+4)(b^2+8)\right)e^{\frac{b^2}{16}}\sqrt{\pi} \right).$$

However, carrying out a numerical integration shows that the expression above is manifestly erroneous, as it does not render the exact kinetic energy value, contrary to the result of eq. (19) in ref. [8]. In addition, the momentum density as defined above does not agree with the analytic expansion of the momentum density as proposed in ref. [9].

While the error in the above expression merely consists of typos, as will appear below, it seems useful to reestablish a correct integral in the present case, at least for the reason that follows. Indeed, although the authors have not mentioned it, the integration method they suggest, invoking elliptic coordinates, and their subsequent result for the momentum density, makes it likely that a closed-form result can be found for the reciprocal form factor, which result has seemingly not been reported to date.

In this respect, the "spinless" reduced density-matrix allows for determining the momentum density, generally defined as

$$n(\mathbf{p}) = \frac{1}{(2\pi)^3}\int \rho(\mathbf{r},\mathbf{r'})\,e^{i\mathbf{p}\cdot(\mathbf{r}-\mathbf{r'})}d\mathbf{r}d\mathbf{r'}.$$

A rotation of the position-space coordinates allows for instance for rewriting it

$$n(\mathbf{p}) = \frac{1}{(2\pi)^3}\int \tilde{\rho}(\mathbf{R},\mathbf{u})\,e^{i\mathbf{p}\cdot\mathbf{u}}d\mathbf{R}d\mathbf{u}, \qquad (2)$$

wherein $\mathbf{R} = (\mathbf{r}+\mathbf{r'})/2$, $\mathbf{u} = \mathbf{r}-\mathbf{r'}$, and $\tilde{\rho}(\mathbf{R},\mathbf{u}) \equiv \rho(\mathbf{r},\mathbf{r'})$.

Now, the so-called reciprocal form factor (or "auto correlation function") $B(\mathbf{u})$ is usually defined as:

$$B(\mathbf{u}) = \int \rho(\mathbf{r}+\mathbf{u},\mathbf{r})\,d\mathbf{r}$$

Yet, $B(\mathbf{u})$ can equivalently be defined as

$$B(\mathbf{u}) = \int \rho(\mathbf{r}+\mathbf{u}/2,\mathbf{r}-\mathbf{u}/2)\,d\mathbf{r}. \tag{3}$$

Thus, $B(\mathbf{u})$ averages out the local information in $\tilde{\rho}(\mathbf{R},\mathbf{u}) \equiv \rho(\mathbf{r},\mathbf{r}')$; in other words, $B(\mathbf{u})$ is the reduced density-matrix projected on the off-diagonal axis defined by $\mathbf{u} = \mathbf{r} - \mathbf{r}'$. This function is therefore of importance inasmuch as it reflects the off-diagonal side of the exact reduced density-matrix. In practice, $B(\mathbf{u})$ is a useful intermediate between the density-matrix expressed in position space and momentum properties [9].

Next, it is implicit from eqs. (2) and (3) that $n(\mathbf{p})$ and $B(\mathbf{u})$ are 3D Fourier transforms of one another. In particular, in case of spherical two-electron systems:

$$\begin{aligned}n(\mathbf{p}) &= \frac{1}{(2\pi)^3}\int B(\mathbf{u})\,e^{i\mathbf{p}\cdot\mathbf{u}}\,d\mathbf{u} \\ &= \frac{1}{(2\pi)^3}\int B(u)\,4\pi u\,\frac{\sin pu}{p}\,du\end{aligned}. \tag{4}$$

In the case of the Hooke's atom, the exact wavefunction leads to the following formal expression of the density-matrix [3,8, 9]:

$$\rho(\mathbf{r}_1,\mathbf{r}_1') = 2\mathcal{N}^2\,e^{-(r_1^2+r_1'^2)/4}\int e^{-r_2^2/2}(1+r_{12}/2)(1+r_{1'2}/2)\,d\mathbf{r}_2, \tag{5}$$

or, in terms of rotated coordinates $\mathbf{R} = (\mathbf{r}+\mathbf{r}')/2$ and $(\mathbf{u}=\mathbf{r}-\mathbf{r}')$

$$\tilde{\rho}_1(\mathbf{R},\mathbf{u}) = 2\mathcal{N}^2\,e^{-(R^2+u^2/4)/2}\int e^{-r_2^2/2}\bigl(1+|\mathbf{r}_2-\mathbf{R}-\mathbf{u}/2|/2\bigr)\bigl(1+|\mathbf{r}_2-\mathbf{R}+\mathbf{u}/2|/2\bigr)\,d\mathbf{r}_2. \tag{6}$$

Introducing now the coordinate $\boldsymbol{\sigma} = \mathbf{r}_2 - \mathbf{R}$, that is, following ref. [8], the above equation somewhat simplifies to:

$$\tilde{\rho}_1(\mathbf{R},\mathbf{u}) = 2\mathcal{N}^2 \, e^{-(R^2+u^2/4)/2} \int e^{-(\sigma+\mathbf{R})^2/2}\left(1+|\sigma-\mathbf{u}/2|/2\right)\left(1+|\sigma+\mathbf{u}/2|/2\right)d\sigma. \tag{7}$$

Thus, using the alternative definition of eq. (3), the reciprocal form factor can be written as

$$B(\mathbf{u}) = 2\mathcal{N}^2 \, e^{-u^2/8} \int\int d\sigma \, d\mathbf{R} \, e^{-R^2/2} e^{-(\sigma+\mathbf{R})^2/2}\left(1+|\sigma-\mathbf{u}/2|/2\right)\left(1+|\sigma+\mathbf{u}/2|/2\right). \tag{8}$$

Now, integrating first over variable **R** leads to:

$$B(\mathbf{u}) = 2 \frac{1}{16(8+5\sqrt{\pi})\pi} e^{-u^2/8} \int d\sigma \, e^{-\sigma^2/4}\left(2+|\sigma-\mathbf{u}/2|\right)\left(2+|\sigma+\mathbf{u}/2|\right), \tag{9}$$

it being understood that the normalization condition imposes $B(0) = 2$ here.

Finally, the above integral can be carried out using elliptic coordinates, as suggested by the authors in [8] as to the calculation of the momentum density. In particular, making use of $\mu = (|\sigma + \mathbf{u}/2|+|\sigma - \mathbf{u}/2|)/s$ and $\nu = (|\sigma + \mathbf{u}/2|-|\sigma - \mathbf{u}/2|)/s$, leads to $B(u) =$

$$\int_1^\infty d\mu \int_{-1}^1 d\nu \, 2\pi \left(\frac{u^3}{8}(\mu^2-\nu^2)\right) 2 \frac{1}{16(8+5\sqrt{\pi})\pi^{5/2}} e^{-\frac{u^2}{8}-\frac{1}{4}\left(\frac{u^2}{4}(\mu^2+\nu^2-1)\right)}$$
$$\pi^{3/2}(2+(\mu+\nu)u/2)(2+(\mu-\nu)u/2)$$

which evaluates in closed-form to

$$B(u) = \frac{1}{4(8+5\sqrt{\pi})u} e^{-\frac{3u^2}{16}} \tag{10}$$

$$\left(8u(4+u) - 64 e^{\frac{u^2}{8}} \pi \left(-1+\mathrm{erf}\left[\frac{u}{4}\right]\right)\mathrm{erf}\left[\frac{u}{4}\right] + \right.$$

$$\left. e^{\frac{u^2}{16}} \sqrt{\pi} \left(-u(-8+u^2) + 2(4+u)(8+u^2)\mathrm{erf}\left[\frac{u}{4}\right]\right)\right),$$

Thus, according to eq. (2), and choosing $u$ as a dummy integration variable ($u \equiv b$ in ref. [8], the integral form of the momentum density could be written as

$$n(p) = \frac{1}{(2\pi)^3} \rho(p) =$$

$$\frac{1}{(2\pi)^3} \frac{\pi}{8+5\sqrt{\pi}}$$

$$\int_0^\infty db \frac{\sin(p\,b)}{p} e^{-\frac{3b^2}{16}}$$

$$\left( 8b(4+b) - 64 e^{\frac{b^2}{8}} \pi \left(-1 + \mathrm{erf}\left[\frac{b}{4}\right]\right) \mathrm{erf}\left[\frac{b}{4}\right] + \right.$$

$$\left. e^{\frac{b^2}{16}} \sqrt{\pi} \left(-b(-8+b^2) + 2(4+b)(8+b^2)\mathrm{erf}\left[\frac{b}{4}\right]\right) \right),$$

wherein $\rho(p)$ denotes the momentum density normalized as in ref.[8]. This expression differs from that of eq. (19) in ref. [8], due to the term erf[$b/4$] in the last member and to the second prefactor, independently of the normalization choice.

Finally, one can verify from eq. (10) above that $B(0) = 2$ and that $-3/2 B''(0) =$

$$-\frac{3}{2}\left(-\frac{7}{15} + \frac{2}{5(8+5\sqrt{\pi})}\right) = 0.6644176,$$

that is, the exact value of the kinetic energy [3].

Accordingly and aside from typos, the result of the authors implicitly points at a way of achieving an exact closed-form expression for the reciprocal form factor of the Hooke's atom.